\documentclass[twoside]{article}
\usepackage{fleqn,espcrc2}

\def\a{\alpha}\def\b{\beta}\def\d{\delta}\def\e{\epsilon}

\newcommand{\AmS}{{\protect\the\textfont2
  A\kern-.1667em\lower.5ex\hbox{M}\kern-.125emS}}

\title{Global Space-time Symmetries Gauging and
Kaluza-Klein Dimensional Reduction}

\author{A. J. Nurmagambetov\address{Institute for Theoretical Physics NSC
``Kharkov Institute of Physics and Technology",\\
Akademicheskaya St. 1, Kharkov, 61108, Ukraine}
\thanks{Permanent Position}
\address{
Dipartimento di Fisica ``Galileo Galilei",
Universit\'a di Padova\\
via F. Marzolo 8, Padova, 35131, Italia}
\thanks{Postdoctoral Fellow}
}

\begin{document}

\begin{abstract}
A relation between dimensional reduction and space-time symmetry gauging
is outlined.
\vspace{1pc}
\end{abstract}

\maketitle

\section{Foreword}

One of the miracles of String theory is its higher dimensional nature. But
since everyone can directly feel and observe only four-dimensional world, it 
is a
great puzzle how to connect the String/M-theory living at a Planck energy 
scale in
ten/eleven space-time dimensions with phenomenological models describing
experimental data. A way to resolve this problem is to
compactify additional dimensions.

In this volume in memory of the 75th anniversary of D.~V.~Volkov
it is worth to emphasize that Dmitrij Vasilievich Volkov was
one of the pioneers in the investigation of mechanisms of spontaneous
compactifications of Kaluza-Klein (KK) theories \cite{cs}--\cite{vt1}. In
\cite{vt}, \cite{vt1} D. V. Volkov and V. I. Tkach proposed for the first 
time the mechanism of spontaneous compactification based on embedding the spin 
connection of the compactified manifold into a gauge field connection, which
was then applied in \cite{vst0}--\cite{st} for studying KK supergravities.

To compactify extra dimensions one needs as a starting
point an {\it ansatz} being the classical solution for the low energy
effective superstring theory (supergravity) equations of motion \cite{dnp},
and the choice of every ansatz compatible with the equations of motion 
determines its own independent classical vacuum state. Small fluctuations
over the vacuum form a spectrum of massless and massive modes with masses 
depending
on internal properties of the compactified manifold. Because there is
no additional rule of the selection of the  ground state metric and 
antisymmetric
tensor fields (for instance, classical stability of vacuum and compactified
manifold can be broken by quantum corrections and non-perturbative
effects)
there exists a problem of the choice of the true classical vacuum state.
However, taking into account the requirement of unbroken supersymmetry one
can restrict the consideration to the ans{\"a}tze which correspond to 
backgrounds which are maximally symmetric from the lower-dimensional observer
point of view, i.e. which are invariant under the space-time group/supergroup
of rotations and translations generated by a set of Killing vectors/spinors.
The form of the maximally-symmetric ground-state ansatz is
governed by the properties of Killing vectors and in general involves a
``warp" factor, which 
is a function of internal coordinates in the Randall-Sundrum
(RS1)/Rubakov-Shaposhnikov (RS2) scenarios \cite{rs}, \cite{rsh}.

This restriction leads to the space-time
manifolds of a constant curvature or Einstein spaces among which only
Minkowski and anti-de Sitter spaces are selected as the spaces compatible
with supersymmetry. In its turn the internal manifold may also possesses the
invariance under the action of an isometry group, which becomes an
internal group of the lower-dimensional theory after compactification and
which is defined by another set of Killing vectors.  The latter play an
important r{\^ o}le in the definition of the massless states in the 
{\it complete}
Kaluza-Klein ansatz especially in the case of dimensional reduction
when all massive modes are ignored and the massless sector of the
compactified theory is considered independently
\footnote{Since all massive modes arising in
the standard KK compactification have an order of the Planck mass they
are not important for low-energy phenomenology. Note however that this is
not true for quantum KK theories and also for RS1/2 scenarios. 
}.  
The form of the KK {\it ansatz for dimensional reduction} is further
restricted by the requirement of its consistency with higher-dimensional
equations of motion. This poses an additional problem of finding a 
consistent ansatz in the case of nontrivial (and nonlinear) reduction.

At this point it is necessary to recall that the  above arguments do
not appeal to the existence of Killing vectors in the original {\it
arbitrary} higher-dimensional background. However, things change if from
the beginning a higher-dimensional background allows for the existence of
an isometry group. This feature is important for the construction of
T(arget-space)-dual theories \cite{rv} as well as for the investigation of
properties of ``non-standard" BPS objects such as KK-monopoles 
\cite{KKmonopole} and space-time-filling branes (see, for instance, 
\cite{sato} and Refs. therein). 
In this case the space-time isometry group is
generated by a set of the space-time Killing vectors and the requirement
of local invariance under isometry group transformations leads to the
introduction of compensator fields akin to that of the KK theories.
The goal of this note is to find the conditions for the
application of the global space-time symmetry gauging technique to the
description of dimensionally reduced theories and to establish a way to
recover dimensionally reduced theories starting from geometrical
properties of manifolds with isometries but without appealing to a
{\it dimensional reduction ansatz}.

To this end we begin
with the consideration of the dimension reduction in 
Kaluza-Klein theories, after that the
philosophy of gauged $\sigma$-models will be discussed. Having these
thesis and anti-thesis the syn-thesis is realized in
establishing a bridge between these two techniques. We conclude by
discussing the results and open questions.

\section{Dimensional reduction in Kaluza-Klein theories}

Because the dimensional reduction in Kaluza-Klein theories is a 
sufficiently well-known subject, we only remind what kind
of geometrical objects appears in this scheme \cite{dnp}. 

In the Kaluza-Klein approach dealing with theories in space-time dimensions 
higher than four
it is assumed that the higher-dimensional space-time
manifold $M^D$ is the direct product of an internal manifold $M^{k}$ and a
space-time manifold $M^{D-k}$. If a group $G$ is the isometry group of the
metric tensor in $M^k$, then $G$ is generated by the set of the Killing vectors
$k^A_i$ $(A=1,\dots, dim~G)$ having the following properties:
$$
{\cal L}_k g_{ij}=0 \Longleftrightarrow \nabla_{(i} k^A_{j)}
\equiv 2\partial_{(i} k^A_{j)}-2\Gamma^l_{~ij} k^A_l
=0,
$$
$$
[k^A,k^B]=f^{AB}_{~~~C}k^C,\qquad k^A=k^{iA}{\partial\over \partial Z^i},
$$
where ${\cal L}_k$ is the Lie derivative in the Killing vector direction,
$g_{ij}$ is the metric tensor in the internal manifold $M^k$
parameterized by the coordinates $Z^i$, $\nabla$ is the covariant
derivative with the standard definition of the Christoffel connection
$$
\Gamma^l_{~ij}={1\over 2} g^{lk}(\partial_j g_{ki}+\partial_i g_{kj}
-\partial_k g_{ij}),
$$
and
$f^{AB}_{~~~C}$ are the structure constants of the isometry group $G$.

The $D$-dimensional metric tensor can be split on to $M^k$ and $M^{D-k}$
blocks as follows 
\footnote{Throughout the paper all hatted quantities indicate an object in
$D$-dimensional space-time with $\hat{X}^{\underline{\hat m}}$
coordinates, unhatted quantities belong to the $(D-k)$-dimensional space-time
with $X^{\underline m}$ coordinates, and the small indices from the middle
of the Latin alphabet are the indices of the internal space parameterized by
$Z^i$.}
$$
{\hat g}_{\underline{\hat m}\underline{\hat n}}(X,Z)=
\left(\begin{array}{cc}
{g}_{\underline{m}\underline{n}}&
{g}_{\underline{m}i}\\
{g}_{i\underline{n}}& {g}_{ij}
\end{array}\right).
$$
To dimensionally reduce the theory it is necessary to choose an {\it ansatz}, 
which shall
define the form of the matrix blocks and will be consistent with
equations of motion for the original $D$-dimensional theory. It is
known that, in particular, the following choice of the off-diagonal
matrix element
$$
{g}_{\underline{m}i}(X,Z)=A^A_{\underline{m}}(X)k^A_i(Z)
+~massive~~modes
$$
is self-consistent (at least in the case of toroidal compactification).
Here a new vector field $A^A_{\underline{m}}$ as a massless mode
appears.  Because we deal with dimensional reduction,
all massive modes are ignored. The complete form of the Kaluza-Klein
ansatz can be extracted from the expression for the line element written
as
$$
d{\hat s}^2=dX^{\underline{m}}\otimes
dX^{\underline{n}}g_{\underline{m} \underline{n}}
-(dZ^i-A^A k^{Ai})
$$
$$
\otimes
(dZ^j-A^A k^{Aj}) g_{ij}.
$$

Under the $D$-dimensional general coordinate transformations
${\hat X}^{\underline{\hat m}}~\hookrightarrow~{\hat X}^{\underline{\hat
m}} -{\hat \xi}^{\underline{\hat m}}({\hat X})$ the metric tensor
transforms as
$$
\d {\hat g}_{\underline{\hat m}\underline{\hat n}}=2\nabla_{(
{\underline{\hat m}}}{\hat \xi}_{\underline{\hat n})}.
$$
For a special choice of the local parameter 
$$
{\hat \xi}^{\underline{\hat m}}(X,Z)=(0,~\e^A(X)k^{Ai}),
$$
which corresponds to local coordinate shifts in the internal manifold
coordinate directions
a transformation rule for the vector field $A^A_{\underline{m}}$ has the
form of the gauge transformation for a Yang-Mills gauge field
with the gauge group $G$
$$
\d A^A_{\underline{m}}(X)=\partial_{\underline{m}}\e^A (X)-
f^A_{~BC}A^B_{\underline{m}}\e^C.
$$
This observation is one of the key points of the Kaluza-Klein philosophy:
The gauge group in $D-k$ space-time dimensions is connected to the isometry
group of the extra dimensions and is a subgroup of the $D$-dimensional
general coordinate transformation (g.c.t.) group \cite{dwitt}--\cite{salam}.

\section{Global space-time symmetries and their gauging}

Let us outline now a resemblance of the construction 
above to that of gauged sigma-model formulation. To this end
consider a $\sigma$-model ($p$-brane) described by the action 
\cite{howe&tucker}--\cite{polyakov}
$$
S=\int d^{p+1}\xi [\sqrt{- h}(\partial_m {\hat X}^
{\underline{\hat m}}
{\hat g}_{\underline{\hat m}\underline{\hat n}}\partial_n
{\hat X}^{\underline{\hat n}}h^{mn} +(p-1))
$$
$$
+\varepsilon^{{m}_1\dots {m}_{p+1}}
{\hat B}_{\underline{\hat m}_1\dots\underline{\hat m}_{p+1}}
\partial_{m_1} {\hat X}^{\underline{\hat m}_1}\dots \partial_{m_{p+1}}
{\hat X}^{\underline{\hat m}_{p+1}}]
$$
with $h_{mn}$ being the metric on the $p$-brane worldvolume.

For special backgrounds allowing for the existence of isometry directions
this action is invariant under the following {\it global} space-time
transformations \cite{bst}--\cite{hull&spence} 
$$
\d {\hat X}^{\underline{\hat m}}=\e^A k^{\underline{\hat m}A}
$$
with a constant parameter $\e^A$. Here $k^{\underline{\hat m}A}$ are the 
Killing vectors generating the isometry group of the background, i.e.
$$
{\cal L}_k~{\hat g}_{\underline{\hat m}\underline{\hat n}}=0,
$$
and the target space antisymmetric gauge field ${\hat B}$ should satisfy the
equation
$$
{\cal L}_k~(d{\hat B})_{\underline{\hat m}_1\dots
\underline{\hat m}_{p+1}}=0.
$$

For further consideration it is important that
having the {\it global} target-space invariance we can gauge the
isometry group
\footnote{Other aspects of space-time symmetry gauging can be found in
\cite{jjmo} (and Refs. therein). From the modern point of view  
the existence of a global symmetry in the original theory is a sufficient
but not necessary (see, for instance, Ref. \cite{bm}) 
condition for a possibility of constructing its dual \cite{quevedo}
.}
, i.e. to require the invariance
under the {\it local} space-time transformations
$$
\d {\hat X}^{\underline{\hat m}}=\e^A({\hat X})~ k^{\underline{\hat m}A}.
$$
To keep the action invariance we have to introduce a gauge field $A^A_m$
which is minimally coupled to the $\sigma$-model and transforming as
$$
\d A^A_m({\hat X})=\partial_m \e^A-f^A_{~~BC}A^B_m \e^C,
$$
which enters the derivative covariant with respect to the local space-time
transformations 
$$
{\sf D}_m {\hat X}^{\underline{\hat m}}=\partial_m {\hat
X}^{\underline{\hat m}}-A^A_m k^{\underline{\hat m}A}.
$$

Gauged $\sigma$-model action can be obtained after replacing the usual
derivative with the covariant one. For keeping the minimal coupling
to the $A^A_m$ we should require \cite{jjmo}, \cite{hull&spence}
$$
{\cal L}_k {\hat B}_{\underline{\hat m}_1\dots\underline{\hat m}_{p+1}}=0.
$$
Therefore
$$
S_{g.}=
\int d^{p+1}\xi [\sqrt{- h}({\sf D}_m {\hat X}^{\underline{\hat
m}} {\hat g}_{\underline{\hat m}\underline{\hat n}}{\sf D}_n
{\hat X}^{\underline{\hat n}}h^{mn} +(p-1))
$$
$$
+\varepsilon^{{m}_1\dots
{m}_{p+1}}{\hat B}_{\underline{\hat m}_1\dots\underline{\hat
m}_{p+1}} {\sf D}_{m_1} {\hat X}^{\underline{\hat m}_1}\dots {\sf
D}_{m_{p+1}} {\hat X}^{\underline{\hat m}_{p+1}}].
$$

\section{The bridge}

We have observed that one of the essential ingredients of
the abovementioned approaches is the notion of the Killing vector. The
properties of dimensional reduction and global space-time symmetry gauging
based on the features of Killing vectors are summarized in the
Table \ref{table:1}.

\begin{table*}[htb]
\caption{Properties of different approaches.}
\label{table:1}
\newcommand{\cc}[1]{\multicolumn{1}{c}{#1}}
\renewcommand{\tabcolsep}{2pc} 
\renewcommand{\arraystretch}{1.2} 
\begin{tabular}{@{}ll}
\hline
Dimensional reduction in KK theories & 
Global space-time symmetry gauging\\

\hline
&\\
$M^D=M^{D-k}\otimes M^k$ & $M^D$\\
&\\
$G^{(k)}$ -- isometry group of $M^k$ (a subgroup of g.c.t. group)
& $G^{(D)}$ -- isometry group of $M^D$\\
&\\
$\d {\hat X}^{\underline{\hat m}}=\e^A (X) k^{Ai}(Z)\d^{\underline{\hat
m}}_i$
& $\d {\hat X}^{\underline{\hat m}}=\e^A ({\hat X})
k^{A\underline{\hat m}}({\hat X})$\\
&\\
$\d A^A_{\underline m}(X)=\partial_{\underline m}\e^A -
f^A_{~BC}A^B_{\underline{m}}\e^C$
& $\d A^A_{\underline{\hat m}}({\hat X})=\partial_{\underline{\hat m}}\e^A
- f^A_{~BC}A^B_{\underline{\hat m}}\e^C$\\
&\\
\hline
\end{tabular}
\end{table*}

Therefore, we can fit the global symmetry gauging approach to the
description of dimensionally reduced theories if we apply the gauged
sigma-model-like technique to a subgroup of g.c.t. group and restrict fields
and gauge parameters by the requirement that they do not depend on
several spatial coordinates regarded as the coordinates of an
effective ``internal" manifold.

To realize this observation consider a space-time manifold with
isometries.
For the sake of simplicity we shall consider the manifold with only
one isometry direction. Restrict now the field configuration of a theory by
the requirement
$$
{\cal L}_k~\Phi=0
$$
for any field $\Phi$. It is always
possible to rotate the coordinate system in such a way that the Killing
vector $k^{\underline{\hat m}}$ will be settled along one of the spatial
directions, say $Z$. In this so-called adapted frame $k^{\underline{\hat
m}}=\d^{\underline{\hat m}}_{\underline{Z}}$, and our restriction becomes
$$
{\cal L}_k~\Phi=\partial_{\underline{Z}}\Phi=0, 
$$
which is the standard requirement for the dimensional reduction in
${\underline{Z}}$ direction.

To describe the geometry of an arbitrary Riemannian manifold we have to
introduce a vielbein and connection one-forms which should be
invariant under the local coordinate transformations along the Killing
vector
$$ {\cal L}_k {\hat E}^{\underline{\hat a}}=0, \qquad {\cal L}_k
{\hat \omega}^{\underline{\hat a}\underline{\hat b}}=0.
$$
If in addition we impose the zero torsion condition
$$
{\cal D}{\hat E}^{\underline{\hat a}}=
d{\hat E}^{\underline{\hat a}}-{\hat E}^{\underline{\hat b}}\wedge
{\hat \omega}_{\underline{\hat b}}^{~\underline{\hat a}}
\equiv {\hat T}^{\underline{\hat a}}=0
$$
we can recover the general expression for the vielbeins \cite{dfr}
$$
{\hat E}^{\underline a}=e^{\a\phi}E^{\underline a}, \qquad
{\hat E}^{\underline z}=e^{\b\phi}(dZ-A)
$$
in the Kaluza-Klein ansatz which form triangular matrix
$$
{\hat E}_{\underline{\hat m}}^{~\underline{\hat a}}(X,Z)=
\left(\begin{array}{cc}
{\hat E}_{\underline{m}}^{~\underline{a}}(X)&
{\hat E}_{\underline{m}}^{~\underline{z}}(X)\\
0& {\hat E}_{\underline{Z}}^{~\underline{z}}(X)
\end{array}\right).
$$
It is very important to note that the form of the Kaluza-Klein
ansatz is completely defined by the geometry of the space-time manifold.

We can arrive at
the same result by fitting the gauged $\sigma$-model
technique. To this end note that the requirement of the Kaluza-Klein
scheme to consider the special choice of the general coordinate
transformations -- $\d Z=\e(X)$ in the case under consideration -- is
equivalent to the local space-time transformation
$$
\d {\hat X}^{\underline{\hat m}}=\e(X)k^{\underline{\hat m}}
$$
with the parameter $\e$ which does not depend on $Z$. Then, to require the
invariance of our theory under this transformation we can introduce the
differential
$$
{\sf D}{\hat X}^{\underline{\hat m}}=d{\hat X}^{\underline{\hat m}}
-A k^{\underline{\hat m}},
$$
which is covariant with respect to
the local shift in $Z$ direction when
$$
\d A=d \e(X),
$$
and the covariant vielbein one-form 
$$
{\hat E}^{\underline{\hat a}}({\sf D})\equiv {\sf D}{\hat
X}^{\underline{\hat m}} {\hat E}_{\underline{\hat m}}^{~\underline{\hat
a}}.
$$

An important point is that to establish the connection with the KK approach
the component structure of the vielbein one-form should be 
restricted to the very special diagonal form
of ${\hat E}_{\underline{\hat m}}^{~\underline{\hat a}}(X,Z)$ 
$$
{\hat E}_{\underline{\hat m}}^{~\underline{\hat a}}(X,Z)=
\left(\begin{array}{cc}
{\hat E}_{\underline{m}}^{~\underline{a}}(X)&
0\\
0& {\hat E}_{\underline{Z}}^{~\underline{z}}(X)
\end{array}\right),
$$
which, in particular, means that ${g}_{\underline{z}\underline{m}}=0$. This
kind of representation is similar to that of Ref. \cite{borlaf} where
the construction of a target-space geometry in the presence of isometries
has been discussed.  
In its turn, the transition to the covariant differential deforms this
structure to the triangular Kaluza-Klein form. Therefore, in this 
representation the line element can be simply written as
$$
d{\hat s}^2={\hat E}^{\underline{\hat a}}({\sf D})\otimes
{\hat E}^{\underline{\hat b}}({\sf D}){\hat \eta}_{\underline{\hat
a}\underline{\hat b}}.
$$

For antisymmetric tensor fields the form ${\hat C}^{(n)}({\sf D})$ is
apparently invariant under the local transformation in the Killing vector
direction. Rearranging $n$-form in terms of usual differentials we arrive
at
$$
{\hat C}^{(n)}({\sf D})=C^{\prime (n)}(d)+C^{(n-1)}(d)\wedge dZ,
$$
$$
C^{\prime (n)}=C^{(n)}-C^{(n-1)}\wedge A,
$$
which is nothing but the standard rule of reduction for the $n$-form gauge 
field in the Kaluza-Klein theories.

In particular, since the dynamics of a $p$-brane is described by the
pullback of target-space fields,
we conclude that under assumptions above
{\it
the gauged
$(p+1)$-dimensional $\sigma$-model describes the
dynamics of a
directly dimensionally reduced p-brane}.

This observation is in agreement with a result of Ref. \cite{blo} 
adapted to the massless case.

\section{Discussion and open questions}

To summarize, we have outlined the connection between dimensional reduction
and global space-time symmetries gauging. By presenting the line element 
in a more ``economic'' form by means of ``covariant"
vielbein one-forms and the reduction law for the antisymmetric target-space 
tensor fields, the investigation of the reduced theory symmetry structure
is simplified especially in the case when the original theory has a 
non-trivial symmetry structure. An example of such a theory is the
covariant formulation of the M5-brane
\cite{pst}, \cite{blnpst}, \cite{M5}.
The action for the bosonic M5-brane propagating in a D=11 background has the 
form
$$
S_{M5}=-\int\, d^6\xi~[\sqrt{-\det(\hat{g}_{mn}+i\hat{H}^*_{mn})}
$$
$$
+
{\sqrt{-\det \hat{g}_{mn}}\over 4}\hat{H}^{*mn}\hat{H}_{mn}]
+\int_{{\cal M}^6}~{\cal L}_{WZ}(d).
$$
Here $\hat{g}_{mn}=\partial_m \hat{X}^{\underline{\hat m}}
\hat{g}_{\underline{\hat m}\underline{\hat n}}
\partial_n \hat{X}^{\underline{\hat n}}$ is the induced metric,
$H^{(3)}=db^{(2)}-\hat{C}^{(3)}$ is the curl of the worldvolume
second-rank chiral form and
$$
\hat{H}^*_{mn}={1\over \sqrt{-\partial a \hat{g} \partial a}}
\hat{H}^*_{mnr}\partial^r a \equiv (\hat{H}^* \cdot \hat{v})_{mn}, 
$$
$$
\hat{H}_{mn}=({H} \cdot \hat{v})_{mn}, 
\hat{H}^{*mnl}={1\over 3!\sqrt{-\hat{g}}}\e^{mnlpqr}H_{pqr}.
$$
The explicit form of the Wess-Zumino term is
$$
\int_{{\cal M}^6}~{\cal L}_{WZ}(d)=\int_{{\cal M}^6}(\hat{C}^{(6)}+
{1\over 2}db^{(2)}\wedge \hat{C}^{(3)}).
$$
The details of the symmetry structure of this action can be found in
literature.

Let us now consider an eleven-dimensional background having isometry along the
coordinate $Z$. Then after transition to the target space fields which contain
covariant derivative or differential, i.e.
$$
\hat{g}_{mn} \rightarrow \hat{g}_{mn}({\sf D})={\sf
D}_m\hat{X}^{\underline{\hat m}}\hat{E}_{\underline{\hat
m}}^{~\underline{\hat a}}\hat{\eta}_{\underline{\hat a}\underline{\hat b}}
{\sf D}_n\hat{X}^{\underline{\hat n}}\hat{E}_{\underline{\hat
n}}^{~\underline{\hat b}},
$$
$$
\hat{C}^{(n)}(d)\rightarrow \hat{C}^{(n)}({\sf D})
$$
and leaving the worldvolume fields ($b^{(2)}$ and $a$) the same we get the M5
gauged sigma-model action
$$
S_{M5g.}=-\int\, d^6\xi~[\sqrt{-\det(\hat{g}_{mn}({\sf
D})+i\hat{H}^*_{mn}({\sf D}))}
$$
$$
+
{\sqrt{-\det \hat{g}_{mn}({\sf D})}\over 4}\hat{H}^{*mn}({\sf
D})\hat{H}_{mn}({\sf D})] 
$$
$$
+\int_{{\cal M}^6}~{\cal L}_{WZ}({\sf D}).
$$
In view of the statement of the previous section, this action is nothing
but the action for an NSIIA five-brane \cite{bns}.
Because the structure of the NSIIA five-brane action
formally remains the same it is evident that the symmetry structure of the
NSIIA5-brane is akin to that of the M5-brane. New gauge symmetries arising 
under reduction can be derived from the definition of the covariant derivative.

Consider now the application of the proposed scheme to
the description of dimensionally reduced (super)gravity theories. To this 
end recall that the procedure of dimensional reduction, say,
onto multidimensional torus can be reformulated
in the following way. Imagine that our action functional
is described by a $p$-form integrated over $p$-dimensional manifold ${\cal
M}^p$ being the 
direct product of the space-time manifold ${\cal M}^{p-n}$
and $n$-dimensional torus $T^n$. The latter is parameterized by
$Z^i$ coordinates with $i=1,\dots,n$. Then we can write down the
action as
$$
S=\int_{{\cal M}^p}\, {\cal L}^{(p)}
$$
$$
=
\int_{{\cal M}^{p-n}\times {\cal M}^n}\,
\big [\alpha~ i_{{i_1}}\dots i_{{i_n}}({\cal L}^{(p)}\wedge
dZ^{i_1}\wedge\dots\wedge dZ^{i_n})
$$
$$
+ \omega~ i_{{i_1}}\dots i_{{i_n}}{\cal
L}^{(p)}\wedge dZ^{i_1}\wedge\dots\wedge dZ^{i_n} \big ]
$$
with numerical coefficients $\alpha$ and $\omega$.
The first integral is zero as it should be for a $p$-form on the
$p$-dimensional manifold, which does not contain $n$ differentials
of the manifold coordinates. Upon integration over $Z^i$ the second term
takes the form
$$
S_{Red.}=\Omega_{{\cal M}^n}\int_{{\cal M}^{p-n}}\,
\e^{i_1\dots i_n}
i_{{i_1}}\dots i_{{i_n}}{\cal
L}^{(p)}, 
$$
$$
\Omega_{{\cal M}^n}^{i_1\dots i_n}=\int_{{\cal M}^{n}}\, \omega~
dZ^{i_1}\wedge\dots dZ^{i_n},
$$
where $\Omega_{{\cal M}^n}$ is the global volume of the internal space
which we take to be of a unite value.

In the simplest case of ${\cal M}^p={\cal M}^{p-1}\times
{\cal M}^1$ 
\begin{equation}\label{Lp}
\int_{{\cal M}^p}\,{\cal L}^{(p)}(d)
=\int_{{\cal M}^p}\,
i_Z {\cal L}^{(p)}\wedge dZ.
\end{equation}
On the other hand, by definition
$$
{\cal L}^{(p)}({\sf{D}})={1\over p!}{\sf{D}}X^{\underline{{\hat
m}_1}}\wedge\dots \wedge {\sf{D}}X^{\underline{{\hat m}_p}}{\cal
L}_{\underline{{\hat m}_p}\dots \underline{{\hat m}_1}}
$$
$$
={\cal L}^{(p)}(d)-i_k {\cal L}^{(p)}\wedge A.
$$
Therefore taking into account eq. (1) and choosing the adapted coordinate frame
one arrives at
\begin{equation}\label{LpD1}
{\cal L}^{(p)}({\sf{D}})=
i_Z {\cal L}^{(p)}(d)\wedge
(dZ-A).
\end{equation}
Integrating (2) over $Z$ we recover the result of dimensional
reduction.

To illustrate how it works, consider the classical example of the reduction 
of a five-dimensional Kaluza-Klein gravity down to four dimensions
in a way presented in \cite{dfr}.
The $D=5$ gravity action 
$$
S=\int_{{\cal M}^5}\, {1\over  3!} {\hat R}^{\underline{\hat
a_1}\underline{\hat a_2}}\wedge \hat{E}^{\underline{\hat a_3}} \wedge
\hat{E}^{\underline{\hat a_4}} \wedge \hat{E}^{\underline{\hat a_5}}
\e_{\underline{\hat a_1}\dots \underline{\hat a_5}}
$$
$$
\equiv \int\, d^5 x \, \sqrt{-{\hat g}}{\hat R}.
$$
Consider now the Lagrangian 5-form
$$
{\cal L}^{(5)}({\sf{D}})
={1\over  3!} {\hat R}^{\underline{\hat
a_1}\underline{\hat a_2}}({\sf{D}})\wedge \hat{E}^{\underline{\hat a_3}}
({\sf{D}}) \dots
\hat{E}^{\underline{\hat a_5}}({\sf{D}}) \e_{\underline{\hat a_1}\dots
\underline{\hat a_5}},
$$
which by use of the representation for the curvature and vielbein forms
$$
{\hat R}^{\underline{\hat a}\underline{\hat b}}({\sf{D}})=
{\hat R}^{\underline{\hat a}\underline{\hat b}}(d)-
d A(i_k \hat{\omega}^{\underline{\hat a}\underline{\hat b}})-
A {\cal D}(i_k \hat{\omega}^{\underline{\hat a}\underline{\hat b}}),
$$
$$
{\hat E}^{\underline{\hat a}}({\sf{D}})={\hat E}^{\underline{\hat a}}(d)
-A(i_k {\hat E}^{\underline{\hat a}})
$$
is written as
$$
{\cal L}^{(5)}({\sf{D}})=-
{1\over 3!}[3({\hat R}^{\underline{a_1}\underline{
a_2}}-d A (i_k \hat{\omega}^{\underline{ a_1}\underline{
a_2}})) 
$$
\begin{equation}\label{L5}
\wedge \hat{E}^{\underline{a_3}} \wedge
\hat{E}^{\underline{a_4}} \wedge (dZ-A)
\e_{\underline{a_1}\dots \underline{a_4}}
\end{equation}
$$
-3{\cal D}(i_k \hat{\omega}^{\underline{ a_1}\underline{
a_2}})\wedge \hat{E}^{\underline{ a_3}} \wedge
\hat{E}^{\underline{a_4}}
\wedge A\wedge (dZ-A)
\e_{\underline{a_1}\dots \underline{a_4}}],
$$
where we have used that $i_k \hat{E}^{\underline{\hat a_5}}=
\d^{\underline{\hat a_5}}_{\underline{z}}$ and 
$\e_{\underline{\hat a_1}\dots \underline{\hat a_4}
\underline{z}}=-\e_{\underline{ a_1}\dots \underline{ a_4}}$.

What we can say about the last term of eq. (3)? 
It is not so difficult to observe that
the torsion two-form ${\hat T}^{\underline{a}}$ vanishes. The
Bianchi identity
$
{\cal D}{\hat T}^{\underline{a}}={\hat E}^{\underline{b}}
{\hat R}_{\underline{b}}^{~\underline{ a}}
$
implies in particular
$
i_k {\hat R}_{[\underline{abc}]}=0,
$
or, in view of the requirement ${\cal L}_k {\hat \omega}^{\underline{\hat a}
\underline{\hat b}}=0$,
$$
{\cal D}_{[\underline{a}}i_k {\hat
\omega}_{\underline{b}\underline{c}]}=0.
$$
Hence, the term under consideration vanishes and 
$$
i_k {\hat \omega}_{\underline{a}\underline{b}}=
-{1\over 2}\partial_{[\underline{a}} A_{\underline{b}]}=-{1\over 4}
F_{\underline{a}\underline{b}}.
$$
Integrating over $Z$ and taking into account
$$
\ast F_{\underline{c}\underline{d}}=-{1\over
2}\e^{\underline{a}\underline{b}}_{~~\underline{c}\underline{d}}
F_{\underline{a}\underline{b}}
$$
one arrives at
$$
S_{Red.}=-\int_{{\cal M}^4}\, [{1\over 2!}{\hat R}^{\underline{a_1}
\underline{a_2}}\wedge {\hat E}^{\underline{a_3}}
\wedge {\hat E}^{\underline{a_4}} \e_{\underline{a_1}\dots
\underline{a_4}}
$$
$$
-{1\over 2}F^{(2)}\wedge\ast F^{(2)}].
$$
By rescaling the  vielbeins we can rewrite the action in the Einstein frame 
$$
S_{Red.}=\int\,d^4 x \sqrt{-g}[R-{1\over 2}(\partial \phi)^2
+{1\over 4}e^{-\sqrt{3}\phi}F^2],
$$
which is a well known result (cf. \cite{dnp}).

Thus, in the simplest case it does work. Moreover, it works in the case when
antisymmetric tensor fields enter the action which describes a bosonic
sector of supergravity. It is clear also, that the generalization to the
reduction onto $T^n$ is straightforward and in this case, because the reduction
can be performed in a step-by-step manner, we have the embedding chain, which
we call ``matryoshka'', of the lower-dimensional theories into the 
original one
which should lead to the result similar to that of \cite{pope}.

Several concluding  remarks are in order. Firstly, we may expect
that this scheme can be modified to describe 
more complicated cases of the reduction onto G/H coset manifolds and $S^n$,
although, apparently, the hint with fixing an adapted coordinate frame will
not work.
Secondly, the approach we discuss is restricted to the pure bosonic 
consideration and the application of this scheme has allowed, for 
instance, to make a new test
for T-duality in the bosonic string theory \cite{nur}.

Therefore, the problems for further study are to extend this
approach to the non-Abelian and to the supersymmetric cases. We hope
to make a progress in these directions.

\vspace{0.8cm}

{\bf Acknowledgements}

\vspace{0.3cm}

I would like to acknowledge partial support of the
D. V. Volkov's memorial conference by the INTAS and would like to thank the 
stuff of the Institute for
Theoretical Physics of NSC KIPT for kind help and collaboration in
organizing this meeting. Special thanks to all the participants of this
conference. I am very grateful to Dmitri Sorokin for pleasant and
fruitful discussions of Kaluza-Klein theories, 
for careful reading the manuscript and his clarifying criticism thereof. I 
would also like
to thank Paolo Pasti and Mario Tonin for encouragement, Ergin Sezgin for
interest to this work, Igor Bandos and Vladimir Zima 
for the discussion of the subject of this paper. This work is supported in
part by the Ukrainian Ministry of Science and Education Grant N 2.5.1/52.


\begin{thebibliography}{99}

\bibitem{cs}
E. Cremmer and J. Scherk, Nucl. Phys. B118 (1977) 61.

\bibitem{luciani}
J. F. Luciani, Nucl. Phys. B135 (1978) 111.

\bibitem{fr}
P. G. O. Freund and M. A. Rubin, Phys. Lett. B97 (1980) 233.

\bibitem{englert}
F. Englert, Phys. Lett. B119 (1982) 339.

\bibitem{vt}
D. V. Volkov and V. I. Tkach, JETP Lett. 32 (1980) 681.

\bibitem{vt1}
D. V. Volkov and V. I. Tkach, Theor. Math. Phys. 51 (1982) 427.

\bibitem{vst0}
D. V. Volkov, D. P. Sorokin and V. I. Tkach, JETP Lett. 38 (1983) 481;
ibid. 40 (1984) 1162.

\bibitem{vst011}
D. P. Sorokin and V. I. Tkach, Ukr. Fiz. Zh. 29 (1984) 1605; 
ibid. 29 (1984) 1765.

\bibitem{vst1}
D. P. Sorokin, V. I. Tkach and D. V. Volkov, Phys. Lett. B161 (1985) 301;
Yad. Fiz. 43 (1986) 222; ibid. 43 (1986) 443.

\bibitem{st}
D. P. Sorokin and V. I. Tkach, Fiz. Elem. Chast. Atom. Yadra 18 (1987) 1035.


\bibitem{dnp}
M. J. Duff, B. E. W. Nilsson and C. N. Pope,
Phys. Rep. { 130} (1986) 1.

\bibitem{rs}
L. Randall and R. Sundrum, Phys. Rev. Lett. { 83} (1999) 4690.

\bibitem{rsh}
V. A. Rubakov and M. E. Shaposhnikov, Phys. Lett. { B125} (1983) 136.


\bibitem{rv}
M. Ro{\v c}ek and E. Verlinde, Nucl. Phys. {B373} (1992) 630.

\bibitem{KKmonopole}
E. Bergshoeff, B. Janssen and T. Ortin, Phys. Lett. { B410} (1997) 132.

\bibitem{sato}
T. Sato, On M-9-branes and their dimensional reductions, Proceedings of
the International Conference SSQFT2000, Kharkov, Ukraine, hep-th/0102084.


\bibitem{dwitt}
B. S. De Witt, Dynamical Theory of Groups and Fields, Gordon and Breach
Co., New York, 1965.

\bibitem{kerner}
R. Kerner, Ann. Inst. Henri Poincare, Sect. {A9} (1968) 143.

\bibitem{cho}
Y. M. Cho, J. Math. Phys. {16} (1975) 2029.

\bibitem{cho&freund}
Y. M. Cho and P. G. O. Freund, Phys. Rev. { D12} (1975) 1711.

\bibitem{salam}
A. Salam and J. Strathdee, Ann. Phys. {141}
(1982) 316.

\bibitem{howe&tucker}
P. S. Howe and R. W. Tucker, J. Math. Phys. { 19} (1978) 869, 981.

\bibitem{dirac}
P. A. M. Dirac, Proc. R. Soc. London, Ser. { A268} (1962) 57.

\bibitem{polyakov}
A. M. Polyakov, Phys. Lett. {
B103} (1981) 207.

\bibitem{bst}
E. Bergshoeff, E. Sezgin and P. K. Townsend, Ann. Phys. { 185} (1988) 330. 


\bibitem{jjmo}
I. Jack, D. R. T. Jones, N. Mohammedi and H. Osborn, Nucl. Phys. { B332}
(1990) 359.

\bibitem{hull&spence}
C. M. Hull and B. Spence, Nucl. Phys. {B353} (1991) 379.


\bibitem{bm}
A. Bossard and N. Mohammedi, Nucl. Phys. { B595} (2001) 93.

\bibitem{quevedo}
F. Quevedo, Duality and Global Symmetries, Nucl. Phys. B (Proc. Suppl.)
{61A} (1998) 23.

\bibitem{dfr}
R. d'Auria, P. Fre and T. Regge, Group Manifold approach to gravity and
supergravity, in Supergravity'81, CUP 1982, 421.

\bibitem{borlaf}
J. Borlaf, Nucl. Phys. { B514} (1998) 721;\\ 
E. Alvarez, J. Borlaf, J. H. Leon, Phys. Lett. { B421} (1998) 162.



\bibitem{blo}
E. Bergshoeff, T. Ortin and Y. Lozano, Nucl. Phys. {
B518} (1998) 363.


\bibitem{pst}
P. Pasti, D. Sorokin and M. Tonin, Phys. Lett. {B398} (1997) 41.

\bibitem{blnpst}
I. Bandos, K. Lechner, A. Nurmagambetov, P. Pasti, D. Sorokin and M.
Tonin, Phys. Rev.
Lett. {78} (1997) 4332.

\bibitem{M5}
M. Aganagic, J. Park, C. Popescu and J. H. Schwarz, Nucl. Phys. { B496} 
(1997) 191.

\bibitem{bns}
I. Bandos, A. Nurmagambetov and D. Sorokin,
Nucl. Phys. {B586} (2000) 315.

\bibitem{pope}
H. Lu and C. N. Pope, Nucl. Phys. { B465} (1996) 127.

\bibitem{nur}
A. Nurmagambetov, T-duality in the string theory effective action with a string
source, hep-th/0102109.



\end{thebibliography}
\end{document}